**A Self-Consistent Marginally Stable State for Parallel Ion Cyclotron Waves**

Philip A. Isenberg[1a)]

[1] *Institute for the Study of Earth, Oceans and Space, University of New Hampshire, Durham, New Hampshire 03824, USA*

**[Article in press in Physics of Plasmas]**

We derive an equation whose solutions describe self-consistent states of marginal stability for a proton-electron plasma interacting with parallel-propagating ion cyclotron waves. Ion cyclotron waves propagating through this marginally stable plasma will neither grow nor damp. The dispersion relation of these waves, $\omega(k)$, smoothly rises from the usual MHD behavior at small $|k|$ to reach $\omega = \Omega_p$ as $k \rightarrow \pm\infty$. The proton distribution function has constant phase-space density along the characteristic resonant surfaces defined by this dispersion relation. Our equation contains a free function describing the variation of the proton phase-space density across these surfaces. Taking this free function to be a simple "box function", we obtain specific solutions of the marginally stable state for a range of proton parallel betas. The phase speeds of these waves are larger than those given by the cold plasma dispersion relation, and the characteristic surfaces are more sharply peaked in the $v_\perp$ direction. The threshold anisotropy for generation of ion cyclotron waves is also larger than that given by estimates which assume bi-Maxwellian proton distributions.

## I. INTRODUCTION

The ion cyclotron (IC) wave is the dispersive extension of the shear Alfvén wave and as such is a fundamental mode of a plasma. Small-amplitude IC waves propagating through a plasma will interact with the ions through cyclotron or Landau resonances, leading to efficient energy exchange between the waves and particles. Depending on the structure of the ion distributions, waves will either grow or damp while the particles scatter to states of correspondingly lesser or greater energy. These kinetic interactions are well described by quasilinear theory,[1-3] which provides integral expressions for both the wave propagation through a given medium and the ion response to a given wave mode.[4-6] However, the meaningful

a) Electronic mail: phil.isenberg@unh.edu

application of these expressions to modeling the kinetic evolution of a collisionless plasma has been hampered by the lack of self-consistency between the wave and particle descriptions being used.

In the standard usage, a wave dispersion relation, $\omega(\mathbf{k})$, is derived by assuming a particular structure for the ion distribution function, $f(\upsilon_\parallel, \upsilon_\perp)$. Here, $\mathbf{k}$ is the wavevector, taken to be a real quantity for the purposes of this paper, and $\omega$ is the wave frequency, which can be complex, $\omega = \omega_r + i\gamma$. The ion speeds are designated parallel or perpendicular relative to the large-scale magnetic field, $\mathbf{B}$, where we assume here that the ions are distributed gyrotropically about the field direction. In some models this ion distribution is taken to be cold, with no thermal speeds at all, while in other work a thermal structure with a specific functional form, such as a bi-Maxwellian or kappa function velocity dependence is taken. In these pre-defined thermal plasmas, the imaginary part of the wave frequency is always strongly negative for large $k_\parallel$, corresponding to resonant absorption of the wave power by the thermal core ions. Ion beams or anisotropies can also cause $\gamma > 0$ in an intermediate range of $k_\parallel$, and the complex properties of the wave frequency have been extensively tabulated for a wide variety of ion distributions.

However, in reality the ion distributions do not remain in the assumed form for long, and the derived dispersion relations can become substantially less accurate as time goes on. In a collisionless plasma, such as the solar wind, the resonant interactions cause the ions to diffuse along characteristic paths in phase space, defined in the next section. These ion paths are determined by the cyclotron resonance condition, which itself depends on the wave dispersion relation. The coupled system of waves and particles evolves together, and after several growth or damping times, $t \sim 1/\gamma$, it will no longer be well described by the initial ion distribution or the initial wave dispersion relation. The standard tactic at this point is to turn to plasma simulations to follow the subsequent evolution.[7-13] Such work has been valuable, but can still depend on details of the initial conditions, computational resolution and geometry, numerical schemes to supply the free energy to drive an ongoing interaction, finite duration of the computational run, and ambiguous interpretations of the final plasma properties.

In this paper, we consider a different procedure, based on the quasilinear equations, to obtain self-consistent marginally stable states of the coupled wave-particle system. Any system of waves and particles which exhibits linear wave growth or damping will act to adjust itself in the direction toward such a marginally stable state. Since the linear growth/damping rates are proportional to the ion density gradients along the characteristic paths of the resonant interaction, a marginally stable distribution is structured such that its contours of constant phase-space density are aligned with these paths. The ion cyclotron dispersion relation corresponding to such a distribution will have $\gamma(k) = 0$ for all $k$ by construction, so IC waves will propagate through this plasma without growth or damping. (We note that, following Davidson and Ogdan,[14] the term "marginal stability" is frequently used to designate the isolated frequency and wavenumber at which a generally non-zero $\gamma(k)$ passes through zero. In contrast to this usage, in this paper we employ this term to refer to the entire system of particles and waves.)

We will concentrate here on the case of IC waves propagating parallel or anti-parallel to **B**. There are formal reasons for this restriction which we will address in the next section, but our primary justification is phenomenological. Ion cyclotron waves in the solar wind are observed to be highly field-aligned[15, 16] and their interactions with the thermal plasma are also tend to be maximized when $k_\perp = 0$. Furthermore, Chandran et al.[17] have shown that a proton-electron plasma which is resonantly absorbing obliquely-propagating IC waves will respond by emitting new IC waves which are increasingly aligned with the magnetic field. This quasilinear interaction will eventually result in an effective focusing of oblique IC waves into the parallel direction. Chandran et al. infer that a spectrum of resonant IC waves propagating through a collisionless plasma over a broad range of angles should naturally evolve to a spectrum which is dominated by parallel-propagating wave power.

The self-consistent marginally stable states we will derive have a number of immediate applications in the context of collisionless plasmas. Most trivially perhaps, these marginally stable states represent the time-asymptotic result of a system of steadily driven waves and homogeneous plasma, to which the saturation states of driven plasma simulations may be compared in order to test their accuracy.

More fundamentally, this marginal stability defines the threshold anisotropy for resonant generation of IC waves. Such threshold anisotropies have been recognized as important diagnostics for the protons in the solar wind[18-25] and for collisionless regulation of the ion distributions in the Earth's magnetosheath and outer magnetosphere.[9, 26-31] We define the proton anisotropy, $A = \beta_\perp / \beta_\parallel$, as the ratio of the proton perpendicular thermal energy to the parallel thermal energy, where the proton betas are the usual ratios of the respective proton pressures to the magnetic pressure, $\beta_{\parallel,\perp} = 8\pi P_{\parallel,\perp}/B^2$. The threshold anisotropy, above which the plasma is unstable to the growth of IC waves, can be considered as a function of the proton parallel beta.[32, 33] A reliable threshold anisotropy function can serve as a useful closure condition in a collisionless plasma, allowing an approximate fluid treatment of its behavior.[27, 28, 32]

At present, however, these thresholds have been estimated assuming a pre-defined structure of the proton distribution, usually taken as a bi-Maxwellian in order to utilize the familiar plasma dispersion function. In these calculations, such as those presented by Gary et al.,[32] the proton $\beta_{\parallel}$ is set and the IC dispersion relation is explored for bi-Maxwellians with a range of $\beta_{\perp}$. A large enough anisotropy will yield instability, $\gamma(k) > 0$, for a range of wavenumbers, and maximum growth rates (independent of the wavenumber where they are found) are tabulated. The threshold anisotropy is then considered to be that value which corresponds to a sufficiently small maximum ion cyclotron growth rate (such as $\gamma_{max} = 10^{-3}\,\Omega_p$). However, these bi-Maxwellian distributions all have substantial damping rates at wavenumbers above the unstable range, so are actually far from marginal stability. We will compare the anisotropies of the rigorous marginally stable distributions obtained here to the previous estimates and show that the bi-Maxwellian analysis underestimates the true marginally stable anisotropy.

The use of bi-Maxwellian distributions in this context has been defended by Gary et al.[34] on the basis of hybrid simulation results. They claim that their simulations of this instability on an initially bi-Maxwellian proton distribution "preserve(s) the bi-Maxwellian character" of the distribution while reducing the anisotropy. We point out that the simulations in this work were not driven, but rather described the short-term relaxation of an initially unstable distribution. The approximately saturated state which results from this isolated event may not be relevant to a system being continually driven to threshold. We can think of no physical reason why a bi-Maxwellian distribution would be maintained by a collisionless plasma in the presence of continually dissipating IC waves (as may be occurring in the solar wind) or continual injection of perpendicular ions (as is invoked in the magnetosphere).

Finally, the analysis of this paper can be applied to the characteristic paths themselves, which are used to model the kinetic evolution of proton distributions in a variety of space physics contexts. These paths formally denote "trajectories" in velocity space along which the ions will diffuse through their resonant interaction with IC waves. As such, they have been incorporated into models of the resonant heating of protons and minor ions in the collisionless coronal hole,[35-37] to calculating the turbulent heating of the distant solar wind by interstellar pickup ions,[38, 39] and to deriving velocity-space diffusion coefficients in the magnetosphere.[40-42] These characteristic paths have also been identified with "quasilinear plateaus" in observed solar wind proton distributions, which have been taken as evidence of efficient resonant scattering and heating by these waves.[20, 21, 43]

However, a difficulty encountered in these models is that, in order to treat the evolution of the bulk ion population as opposed to just an energetic tail, the characteristic paths are needed for all relevant values of $v_\parallel$ including the small values within the thermal core of the ion distribution. As we describe in the next section, these characteristic paths require knowledge of the phase speed of the resonant waves, which are found at large $k$ for small $v_\parallel$. However, it is well known[6] that IC waves calculated from bi-Maxwellian particle distributions are all heavily damped within the ion thermal core, so their phase speeds, $\omega_r/k$, are not physically meaningful at large $k$. Furthermore, bi-Maxwellian dispersion relations generally do not even yield a solution to the resonance condition in these highly damped cases. Physically, this simply means that a bi-Maxwellian plasma being heated by absorption of ion cyclotron waves does not remain in the bi-Maxwellian state. Practically, though, one still needs a relevant dispersion relation at high wavenumber to obtain the characteristic paths and time-asymptotic ion distributions. A useful assumption has been to simply take the expressions for a cold plasma, which yield valid phase speeds for all $\mathbf{k}$.[35-37, 40, 41, 44, 45] One suspects that this approximation is reasonable for small enough values of the plasma beta, but this suspicion is difficult to verify. Furthermore, there is an uncomfortable contradiction in that one is describing the resonant ion response using a wave description which assumes no resonant particles. In this paper, we will remove this contradiction and obtain characteristic paths which are self-consistently related to the waves that define them. We will then be able to address the limits of validity of the cold-plasma approximation in these models.

In the next section, we will describe the resonant cyclotron wave-particle interaction and summarize the general behavior of the waves and particles in terms of the dispersion relation and the characteristic paths. In Section 3, we derive the self-consistent dispersion relation for the marginally stable state of parallel ion cyclotron waves in a collisionless proton-electron plasma.

This equation still has a free function that needs to be chosen to obtain numerical solutions. In Section 4, we choose a simple box function and present solutions of this self-consistent marginally stable state. Section 5 contains a discussion of these results and our conclusions.

## II. THE CYCLOTRON RESONANT WAVE-PARTICLE INTERACTION

The coupled system of waves and particles interacting through cyclotron or Landau resonances is well described by the quasilinear formalism.[1, 4-6] The general equations are somewhat complicated, but much insight can be gained from first considering the resonant interaction between a single wave and a single particle.

A plasma wave and particle can interact resonantly, efficiently exchanging energy, when

$$\omega(\mathbf{k}) - k_\parallel \upsilon_\parallel = n\Omega_i \tag{1}$$

where $\Omega_i = q_i B / m_i c$ is the gyrofrequency of the particle of species $i$, charge $q_i$, and mass $m_i$, $c$ is the speed of light, and $n$ is any integer. The cyclotron resonances, $n \neq 0$, occur when the wave frequency Doppler-shifted by the particle's motion along the field is matched by a multiple of the particle's gyrofrequency. The Landau resonance at $n = 0$ occurs when the parallel phase speed of the wave, $\omega / k_\parallel$, is equal to the parallel particle speed. The gyrophase-averaged response of the particle conserves the particle energy in the reference frame moving with the parallel phase speed of the resonant wave. The resonant particle thus follows a path given by[44, 46, 47]

$$\frac{d\upsilon_\perp}{d\upsilon_\parallel} = \frac{V_{ph} - \upsilon_\parallel}{\upsilon_\perp}. \tag{2}$$

Here, we define the function $V_{ph}(\upsilon_\parallel)$ as the parallel phase speed of a wave which is resonant with the particle moving at $\upsilon_\parallel$. This function depends on both the $q/m$ of the particle and the dispersion relation of the wave.

This phase-speed function is important for the analysis to follow. Figure 1 provides a graphical illustration of this function for the case of protons in fundamental resonance ($n = 1$) with cold-plasma, parallel-propagating IC waves. We show the dispersion curve $\omega(k)$ for these waves in black. The particular wave mode resonating with a proton of parallel speed $\upsilon_\parallel$, though (1), is given by the intersection of this curve with a straight line of slope $\upsilon_\parallel$ and intercept $(\omega, k) = (\Omega_p, 0)$ (red dashed line). The phase-speed function $V_{ph}(\upsilon_\parallel)$ is then defined as the slope of the line from the origin to the intersection point (blue dashed line), as the proton parallel speed is varied. It depends on the dispersion relation, but is not directly a function of $\omega$ or $k$.

The integral of (2) defines a set of surfaces in phase space, given by

$$\eta^2 \equiv \upsilon_\perp{}^2 + \upsilon_\parallel{}^2 - 2\int^{\upsilon_\parallel} V_{ph}(\upsilon_\parallel{}')d\upsilon_\parallel{}' = \text{const.}, \tag{3}$$

which describe the resonant characteristic paths for a given wave dispersion relation. These resonant surfaces are labeled by their value of $\eta$, which can be visualized as the value of $\upsilon_\perp$ for a given surface when $\upsilon_\parallel = 0$. Particles interacting with a continuous, randomly-phased spectrum of resonant waves will diffuse in phase space along these resonant surfaces. Thus, the overall evolution of a distribution, $f(\upsilon_\parallel, \upsilon_\perp)$, of resonant particles will be to scatter down whatever density gradients are present along these surfaces. The corresponding growth or damping of the resonant waves can be determined from the net loss or gain of particle energy resulting from this diffusion.

When a variety of resonant wave modes are present, each with their own dispersion relation and phase-space function, the particles will diffuse according to the cumulative scattering due to each available resonance. In these cases, the various resonant surfaces generally intersect within large regions of phase space, and the utility of describing the particle evolution by transport along a given surface is reduced. Of particular relevance here is that if a wave-particle system contains intersecting resonant surfaces which allow a path through phase space to infinite particle energy, that system will not exhibit rigorous marginal stability.[1] In this case, there is always a way for the particles to resonantly gain more energy from the waves. This is the formal reason, alluded to in the Introduction, for our restriction to parallel-propagating IC waves, whose resonant surfaces are nested and do not intersect. Obliquely-propagating ion cyclotron waves permit higher-order resonances ($|n| > 1$ in equation (1)), whose resonant surfaces do intersect each other and there is no rigorous marginal stability. Furthermore, a system containing IC waves propagating within a range of angles with respect to the magnetic field will yield intersecting surfaces even if only the fundamental ($n = 1$) resonance is considered. It is these intersections which lead to the effective focusing of oblique waves into the parallel direction reported by Chandran et al.[17]

Therefore, in this paper we consider only parallel-propagating IC waves. The only cyclotron resonance for these parallel waves is the fundamental $n = 1$, which requires the resonant waves and protons to be related by $k\,\upsilon_\parallel < 0$. Thus, positively-propagating IC waves having $k > 0$ only resonate with protons in the $\upsilon_\parallel < 0$ half of the distribution, and backward $k < 0$ waves interact only with the $\upsilon_\parallel > 0$ protons. Figure 1 illustrates the $k > 0$ resonance condition, while the picture for $k < 0$ can be obtained by inverting the graph across the $k = 0$ line. The fact that protons with a given $\upsilon_\parallel$ have only a single resonant IC wave mode means that the resonant surfaces discussed above cannot intersect, so a marginally stable distribution is possible.

The dispersion relation for parallel-propagating IC waves in a collisionless plasma is well known[4, 6]

$$\omega^2 - c^2 k^2 + \pi\omega \sum_i \omega_{pi}{}^2 \int d\upsilon_\parallel \int \upsilon_\perp{}^2 d\upsilon_\perp \frac{\left(1 - \frac{k\upsilon_\parallel}{\omega}\right)\frac{\partial f_i}{\partial \upsilon_\perp} + \frac{k\upsilon_\perp}{\omega}\frac{\partial f_i}{\partial \upsilon_\parallel}}{\omega - k\upsilon_\parallel - \Omega_i} = 0, \qquad (4)$$

where the plasma frequency of the $i$th species of density $n_i$ is $\omega_{pi}{}^2 = 4\pi\, n_i\, q_i{}^2/m_i$ and $f_i$ is the normalized distribution function of that species. We further simplify our analysis by assuming a plasma consisting of only protons and electrons. We neglect the thermal effects of the electrons by taking them to be cold, and limit the analysis to frequencies $\omega << \Omega_e$. Charge neutrality gives $\omega_{pe}{}^2/\Omega_e = -\,\omega_{pp}{}^2/\Omega_p$, and (4) then reduces to

$$\omega^2 - c^2 k^2 - \frac{\omega\,\Omega_p c^2}{V_A{}^2} + \pi\omega\,\omega_{pp}{}^2 \int d\upsilon_\parallel \int \upsilon_\perp{}^2 d\upsilon_\perp \frac{\left(1 - \frac{k\upsilon_\parallel}{\omega}\right)\frac{\partial f}{\partial \upsilon_\perp} + \frac{k\upsilon_\perp}{\omega}\frac{\partial f}{\partial \upsilon_\parallel}}{\omega - k\upsilon_\parallel - \Omega_p} = 0, \quad (5)$$

where $f(\upsilon_\parallel, \upsilon_\perp)$ now refers to the gyrotropic protons. We take the proton distribution to be symmetric in $\upsilon_\parallel$, which then yields a dispersion relation symmetric in $k$. We will generally discuss $k \geq 0$ in what follows, with the understanding that $\omega\,(-k) = \omega\,(k)$.

Since the Alfvén speed $V_A << c$, we drop the first term in (5) with respect to the others, equivalent to neglecting the displacement current for this analysis. We also switch here to dimensionless variables, where frequencies are normalized to the proton cyclotron frequency $\Omega_p$, wavenumbers are normalized to the proton inertial length $V_A/\Omega_p$, and speeds are normalized to the Alfvén speed $V_A$. In these limits, the dispersion relation for IC waves propagating parallel to the magnetic field through a proton-electron plasma becomes

$$\omega \ + \ k^2 \ - \ \pi\omega \int dv_{\parallel} \int v_{\perp}{}^2 dv_{\perp} \frac{\left(1 - \dfrac{kv_{\parallel}}{\omega}\right)\dfrac{\partial f}{\partial v_{\perp}} + \dfrac{kv_{\perp}}{\omega}\dfrac{\partial f}{\partial v_{\parallel}}}{\omega - kv_{\parallel} - 1} \ = \ 0 \, . \qquad (6)$$

## III.  A STATE OF MARGINAL STABILITY

The general expression for the dispersion relation (6) permits the specification of a self-consistent marginally stable state.  In such a state, the proton distribution will be structured so as to have no density gradients along the relevant resonant surfaces.  At the same time, the wave propagation will be defined by a solution to (6) which has no growth or damping for any value of the wavenumber.  Furthermore, the fact that the cyclotron-resonant interaction between protons and IC waves pushes any such system toward marginal stability means that this proton distribution defines the time-asymptotic steady state for this coupled system of waves and particles.

A proton distribution with no gradients along the resonant surfaces is obtained by insisting that it is a function only of the resonant surface label $\eta$.  Using the definition of $\eta$ in (3), the differential operator inside the integral of (6) collapses to

$$\left(\frac{\omega}{k} - v_{\parallel}\right)\frac{\partial f}{\partial v_{\perp}} + v_{\perp}\frac{\partial f}{\partial v_{\parallel}} = v_{\perp}\left(\frac{\omega}{k} - V_{ph}(v_{\parallel})\right)\frac{1}{\eta}\frac{df}{d\eta} \qquad (7)$$

and the dispersion relation becomes

$$\omega + k^2 - \pi k \int dv_{\parallel} \int v_{\perp}{}^3 dv_{\perp} \frac{\dfrac{\omega}{k} - V_{ph}(v_{\parallel})}{\omega - kv_{\parallel} - 1}\frac{1}{\eta}\frac{\partial f}{\partial \eta} = 0 \, . \qquad (8)$$

At this point, neither the phase-speed function $V_{ph}(v_{\parallel})$ nor the actual dispersion relation $\omega(k)$ are known, but a solution of (8) consistent with the definitions of $V_{ph}$ and $\eta$ will describe a marginally stable state of the plasma.  An immediate consequence is that such solutions will have $\mathrm{Im}(\omega) = 0$ by construction, so the resonant denominator in (8) does not create a singularity in the integral, as can be seen from the definition of $V_{ph}$.  The related fact that a valid, physically reasonable solution for $\mathrm{Re}(\omega)$ exists for all $k$ means that the resonant surfaces can be defined for all proton $v_{\parallel}$, enabling the construction of meaningful diffusion coefficients to describe the plasma evolution.

So far, this construction has not specified the form of $f(\eta)$, and we will need to assume a form to proceed further. For the rest of this paper, we will choose a simple form for this introductory investigation. We will pursue other forms in future studies, but this simple form should contain many of the reasonable physical properties.

## IV. SOLUTION FOR A BOX FUNCTION DISTRIBUTION

To proceed with this initial study, we choose a simple functional form for the dependence of the proton phase-space density on the resonant surface variable $\eta$. We take a "box function" such that the density is constant inside a particular resonant surface, and drops to zero outside that surface. This distribution, symmetric in $v_\parallel$, is expressed as

$$f = \frac{1}{N}S(\Gamma^2 - \eta^2)\,, \qquad (9)$$

where $S(x)$ is the Heaviside step function and $N$ is the total proton density. The parameter $\Gamma$ is the maximum value of $v_\perp$ on the bounding resonant surface. This specification allows the integral over $v_\perp$ to be performed, leaving the dispersion relation as

$$\omega + k^2 + \frac{\pi}{N}\int_{-T}^{T} dv_\parallel \frac{\frac{\omega}{k} - V_{ph}(v_\parallel)}{\frac{\omega-1}{k} - v_\parallel}\left[\Gamma^2 - v_\parallel^2 + 2\int_0^{v_\parallel} V_{ph}(x)dx\right] = 0\,, \qquad (10)$$

where $T$ is the maximum value of $v_\parallel$ on the bounding resonant surface. The definition of $\eta$ yields the relation between $T$ and $\Gamma$

$$T^2 - 2\int_0^T V_{ph}(x)dx = \Gamma^2\,, \qquad (11)$$

which indicates that the quantity in the square brackets of (10) vanishes at the endpoints of the integral. Since protons resonate with IC waves propagating in the opposite direction from their parallel speed ($kv_\parallel < 0$ in equation (1)), we see that $V_{ph}(v_\parallel)$ has a sign opposite to that of $v_\parallel$. Thus, (11) shows that $\Gamma > T$ and the marginally stable distribution is always anisotropic with perpendicular temperature greater than the parallel temperature.

Expressions for various moments of the distribution will be useful in the following discussion, so we tabulate them here, defining a convenient positive quantity

$$K_n = -\int_0^T x^n V_{ph}(x) dx .$$  (12)

The normalizing density, in units of $V_A{}^3$, is then

$$N = \frac{4\pi}{3}\left(T^3 + 3K_1\right).$$  (13)

In these units, the plasma betas of the marginally stable protons in the parallel and perpendicular directions are given by the respective second moments of the distribution, $\beta_\parallel = <\upsilon_\parallel{}^2>$ and $\beta_\perp = <\upsilon_\perp{}^2>/2$. Thus,

$$\beta_\parallel = \frac{4\pi}{15N}\left(T^5 + 5K_3\right)$$  (14)

and

$$\beta_\perp = \frac{2\pi}{N}\left(\frac{2}{15}T^5 + T^2 K_1 + 2K_0 K_1 - \frac{1}{3}K_3 - 2\int_0^T x V_{ph}(x)\int_0^x V_{ph}(y)dydx\right)$$  (15)

The limits of the dispersion relation at low and high wavenumbers are also informative. Taking $\omega, |k| \rightarrow 0$ in (10), and noting that $V_{ph}$ is odd in $\upsilon_\parallel$, we find that

$$\frac{\omega^2}{k^2} = 1 + \frac{2\pi}{N}\left(T^2 K_1 + 2K_0 K_1 - K_3 - 2\int_0^T x V_{ph}(x)\int_0^x V_{ph}(y)dydx\right)$$  (16)

which is recognizable as the MHD phase speed for parallel-propagating waves in an anisotropic plasma

$$\frac{\omega}{k} = \pm V_A \sqrt{1 + \beta_\perp - \beta_\parallel} .$$

In the opposite limit, we take $\omega \rightarrow 1$ as $k \rightarrow \pm\infty$, and assume that the integral in (10) is dominated by the behavior of the integrand near $\upsilon_\parallel = 0$ where $V_{ph}$ is small. Balancing the second and third terms in (10), we find approximately that

$$k^2 \approx \frac{2\pi \Gamma^2}{N}\frac{\omega}{|k|}\int_{-T}^T \frac{d\upsilon_\parallel}{\upsilon_\parallel - \frac{\omega - 1}{|k|}} ,$$  (17)

which leads to

$$\omega(k) \approx 1 - |k|\exp\left(-\alpha|k|^3\right) \quad (k \rightarrow \pm\infty).$$  (18)

We find below that the functional form (18), with $\alpha$ independent of $k$, is verified by the full numerical solution, though the implication from (17) that $\alpha = N/2\pi\Gamma^2$ is not exactly correct. This expression can be inverted through the resonance condition to yield the asymptotic phase-speed function

$$V_{ph}(v_\parallel) \approx -\text{sign}(v_\parallel)\left[-\frac{\alpha}{\ln\left(\left|v_\parallel\right|\right)}\right]^{1/3} \qquad (\left|v_\parallel\right| << 1), \qquad (19)$$

which is seen to rise much more steeply for small $|v_\parallel|$ than the equivalent limit from the cold-plasma dispersion relation, given as[17]

$$V_{ph}(v_\parallel) \approx -\text{sign}(v_\parallel)\left|v_\parallel\right|^{1/3}.$$

The full solutions of (10) are obtained through numerical iteration for various values of $T$. Specifically for each $T$, we choose a fixed set of values for $x \equiv 1/k$ and construct an initial guess for the solution values of $y \equiv 1 - \omega$ at each $x$. The resonance condition then yields values at these points for $v_\parallel = xy$ and $V_{ph}(v_\parallel) = x(1-y)$. With these guessed functions, the integrals in (10) are calculated and the resulting values of the left-hand side are then used to improve the choices for $y(x)$. We continue the iteration until the results for the left-hand side are all $< 10^{-5}$. Table I lists the solutions we have obtained, along with some of the bulk properties of the resulting proton distributions.

These solutions all verify the asymptotic dependence at large $k$, given in (18) and (19). The values of $\alpha$ given in Table 1 are obtained by plotting $\ln[(1-\omega)/k]$ against $k^3$ for each solution and fitting the slope of the high-$k$ portion of the line. We find the function $\alpha(T)$ to be nearly linear in $T$, approximately $\alpha \approx 0.4\ T$. In Figure 2, we plot $\alpha(T)$ (red diamonds), along with the functions $0.4\ T$ (black line) and $N/2\pi\Gamma^2$ as suggested by (17) (dashed blue line).

Figure 3 shows the marginally-stable dispersion curves $\omega(k)$ for four values of $\beta_\parallel$, along with the familiar dispersion relation for the cold plasma, $\omega = k\ V_A\ (1-\omega/\Omega_p)^{1/2}$. We see that IC waves traveling through the marginally stable thermal plasma have larger phase speeds than waves propagating through a cold plasma. These marginally stable solutions extend smoothly all the way to $(\omega, k) = (\Omega_p, \infty)$, in striking contrast to the solutions for bi-Maxwellian plasmas which develop large imaginary parts and generally reduced phase speeds from the effect of thermal particles (see e.g. Yoon et al.[48] for some examples).

Higher wave phase speeds correspond to more anisotropic resonant surfaces, especially for small $\upsilon_\parallel$, as can be seen in Figure 4. Here we plot representative resonant surfaces, defined by (3), for the cases shown in Figure 3. In Figure 4a, the two sets of curves correspond to $\eta =$ 0.5 and 1.0, denoting the place (in units of $V_A$) where they meet the $\upsilon_\perp$-axis. Figure 4b shows two similar sets of curves at $\eta = 0.1$ and 0.2, deeper into the thermal core of a proton distribution. We see that the resonant surfaces obtained from the cold plasma assumption become progressively less accurate for larger $\beta_\parallel$ or for the smaller $\eta$ close to the origin of phase space. One might expect an additional qualitative difference between the surfaces internal and external to the box-function distribution, but this is not particularly evident. (Resonant surfaces outside the distribution are still meaningful in that they control the resonant interaction of nonthermal test protons in the system.) External surfaces will meet the $\upsilon_\parallel$-axis at points greater than the value of $T$ for that case. We see that both green curves and the outer black curve in Figure 4a are external, while all the other solid curves in both parts of the figure are internal. From this small sample, the behavior of these curves seems continuously dependent on $\eta$ and $\beta_\parallel$, without a further distinction. In any case these shapes represent the quasilinear saturation of IC heating, so they are the true "quasilinear plateaus", at least for the case of box-function distributions. It will be interesting to see how they compare to the observed solar wind proton distributions, such as analyzed by Marsch and coworkers.[20, 21, 43]

Finally, the proton distributions derived here represent rigorous marginally stable states, so their anisotropies as listed in Table I can be compared with the frequently used bi-Maxwellian estimates for the threshold anisotropy obtained by Gary et al.[32] These threshold anisotropies are thought to follow an inverse power-law in $\beta_\parallel$,

$$A - 1 = \frac{c}{\beta_\parallel{}^s} \tag{20}$$

where $c$ and $s$ are numerical constants to be fit by theory or simulation results.[33] The published bi-Maxwellian estimates set $s = 0.42$ while $c$ depends on the choice of the maximum growth rate being used: $c = 0.43$ for $\gamma_{max} = 10^{-3}\ \Omega_p$ and $c = 0.35$ for $\gamma_{max} = 10^{-4}\ \Omega_p$. We plot (20) with these parameter values in Figure 5, along with our self-consistent results from Table I, shown as red diamonds. It is clear that the anisotropy of these rigorous marginally stable distributions is larger than the bi-Maxwellian estimates. Furthermore, reducing the $\gamma_{max}$ parameter toward $\gamma_{max} = 0$ takes these estimates in the wrong direction. We fit our anisotropy points to the functional form (20), and find the dependence

$$A - 1 = \frac{0.94}{\beta_\parallel^{0.33}}, \tag{21}$$

shown by the solid black line in Figure 5.

## V. DISCUSSION AND CONCLUSIONS

Kinetic models of plasma evolution through the resonant cyclotron wave-particle interaction require knowledge of the dispersive properties of the resonant waves. These dispersive properties themselves depend on the evolving particle distributions. Thus, an accurate description of driven plasma behavior, such as collisionless IC heating or generation of IC waves through injection of perpendicular ions, should in principle take into account the interplay between the wave and particle properties. Such a driven plasma will eventually adjust itself to a marginally stable state, where any potential wave absorption and ion heating is balanced by equivalent wave generation and ion scattering. These marginally stable states have previously been estimated in non-self-consistent ways, using a cold-plasma dispersion relation[35, 36, 38, 40, 41, 44] or threshold conditions derived from bi-Maxwellian distributions.[28, 32, 42]

The marginally stable dispersion relations and distribution functions derived in this paper are fully self-consistent, and they illustrate the limitations of previous work. The resonant surfaces shown in Figure 4 indicate that thermal protons scattering along the characteristics of the cold-plasma dispersion relation will yield much smaller anisotropies than would result from distributions maintained at self-consistent threshold. This discrepancy becomes larger near the central core of the distribution. Figure 5 shows that these marginally stable states are also more anisotropic than the threshold anisotropies estimated from bi-Maxwellian distributions.

The examples in this paper were obtained for the case of a simple box function $\eta$-dependence of the proton phase-space density. In future work, we will explore more physically reasonable assumptions for this dependence, such as using a Gaussian function. It will also be interesting to see if an equivalent analysis would lead to self-consistent marginally stable states for the cyclotron resonant interaction between electrons and parallel-propagating whistler waves.

**ACKNOWLEDGEMENTS**

The author is grateful for valuable conversations with Sergei Markovskii, Terry Forbes, Ben Chandran, and Bernie Vasquez. This research was funded in part by the NSF Space Weather Program, grant ATM0719738; the NSF SHINE Program, grant AGS 0962506; the NASA Heliophysics Theory Program, NNX11AJ37G; and DoE grant DEFG0207ER46372 to the Center for Integrated Computation and Analysis of Reconnection and Turbulence.

TABLE I.  Marginally stable proton distributions with box-function structure.

| $T$ | $\Gamma$ | $N$ | $\beta_{\parallel}$ | $\beta_{\perp}$ | $\alpha$ | Anisotropy |
|---|---|---|---|---|---|---|
| 0.001 | 0.01308 | $5.912 \times 10^{-7}$ | $1.765 \times 10^{-7}$ | $3.025 \times 10^{-5}$ | $3.635 \times 10^{-4}$ | 171.4 |
| 0.005 | 0.03844 | $2.555 \times 10^{-5}$ | $4.417 \times 10^{-6}$ | $2.613 \times 10^{-4}$ | $1.877 \times 10^{-3}$ | 59.16 |
| 0.01 | 0.06127 | $1.300 \times 10^{-4}$ | $1.769 \times 10^{-5}$ | $6.644 \times 10^{-4}$ | $3.757 \times 10^{-3}$ | 37.57 |
| 0.05 | 0.1831 | $5.840 \times 10^{-3}$ | $4.443 \times 10^{-4}$ | $5.952 \times 10^{-3}$ | $1.879 \times 10^{-2}$ | 13.40 |
| 0.1 | 0.2961 | $3.075 \times 10^{-2}$ | $1.785 \times 10^{-3}$ | $1.562 \times 10^{-2}$ | $3.866 \times 10^{-2}$ | 8.753 |
| 0.2 | 0.4840 | 0.1656 | $7.184 \times 10^{-3}$ | $4.195 \times 10^{-2}$ | $7.683 \times 10^{-2}$ | 5.839 |
| 0.3 | 0.6497 | 0.4506 | $1.624 \times 10^{-2}$ | $7.586 \times 10^{-2}$ | 0.1162 | 4.671 |
| 0.4 | 0.8037 | 0.9245 | $2.899 \times 10^{-2}$ | 0.1165 | 0.1545 | 4.018 |
| 0.8 | 1.365 | 5.419 | 0.1173 | 0.3391 | 0.3209 | 2.890 |
| 1.0 | 1.628 | 9.700 | 0.1841 | 0.4843 | 0.4007 | 2.631 |
| 1.4 | 2.138 | 23.64 | 0.3632 | 0.8400 | 0.5664 | 2.313 |
| 3.0 | 4.084 | 189.3 | 1.694 | 3.112 | 1.237 | 1.837 |

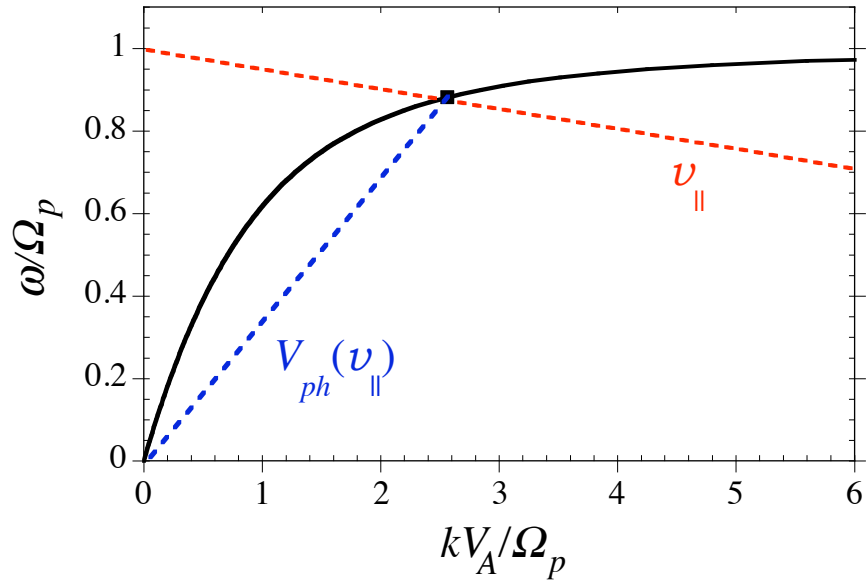

**Figure 1.** Illustration of the cyclotron resonance condition for parallel-propagating IC waves. The intersection of the dispersion curve $\omega(k)$ (black line) with a straight line of slope $\upsilon_{\parallel}$ passing through the point $(\omega, k) = (\Omega_p, 0)$ (red dashed line) indicates the frequency of the resonant wave. The slope of the line between the origin and the resonant point (blue dashed line) defines the resonant phase-speed function $V_{ph}(\upsilon_{\parallel})$.

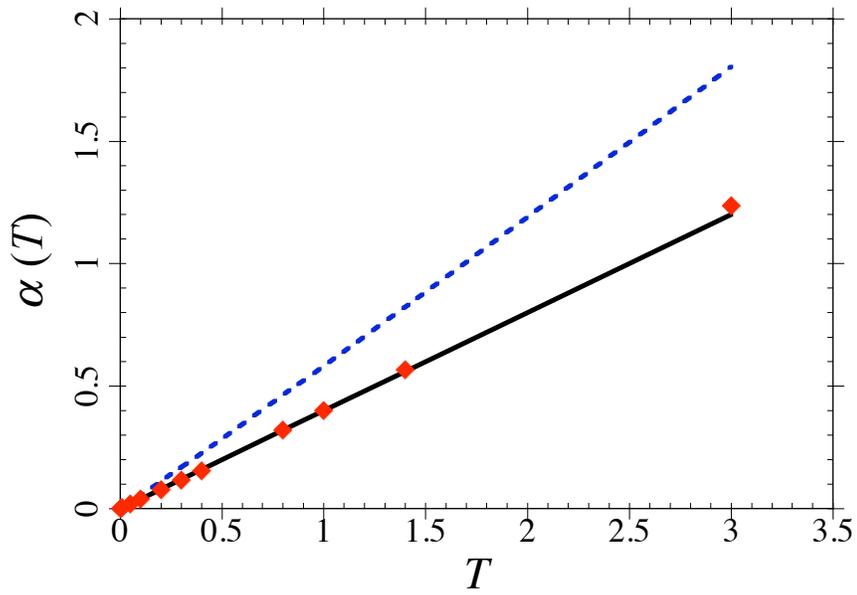

**Figure 2.** Plot of the proportionality term $\alpha\,(T)$ found in the asymptotically high-$k$ expressions (18) and (19). The values from the numerical solutions are shown as red diamonds. For comparison, the black line plots $\alpha = 0.4T$ and the blue line plots $\alpha = N/2\pi\Gamma^2$ as suggested by (17).

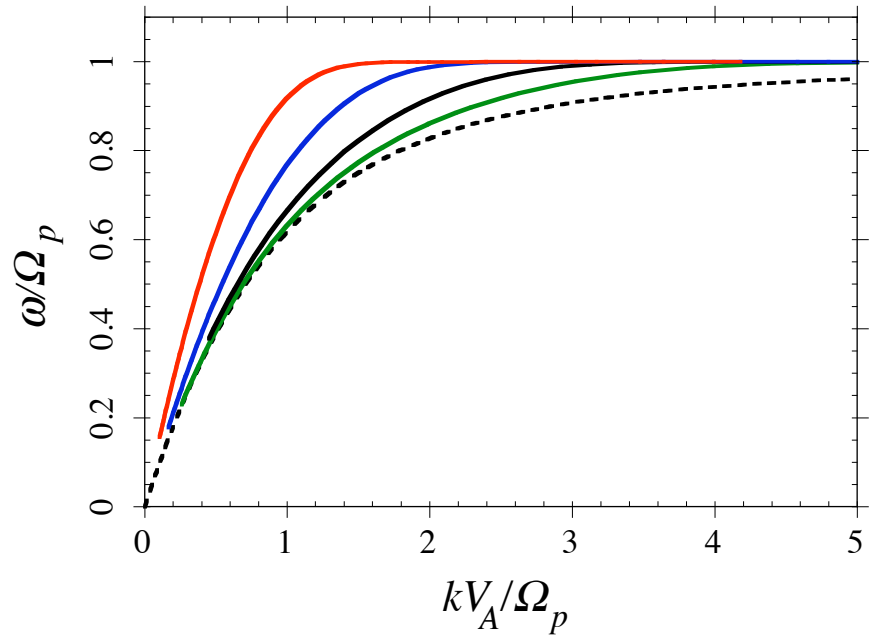

**Figure 3.** Dispersion curves for IC waves in the marginally-stable plasma for four values of $\beta_{\parallel}$: $\beta_{\parallel} = 0.001785$ (green), $\beta_{\parallel} = 0.01624$ (black), $\beta_{\parallel} = 0.1841$ (blue), and $\beta_{\parallel} = 1.694$. The black dashed line shows the cold-plasma solution.

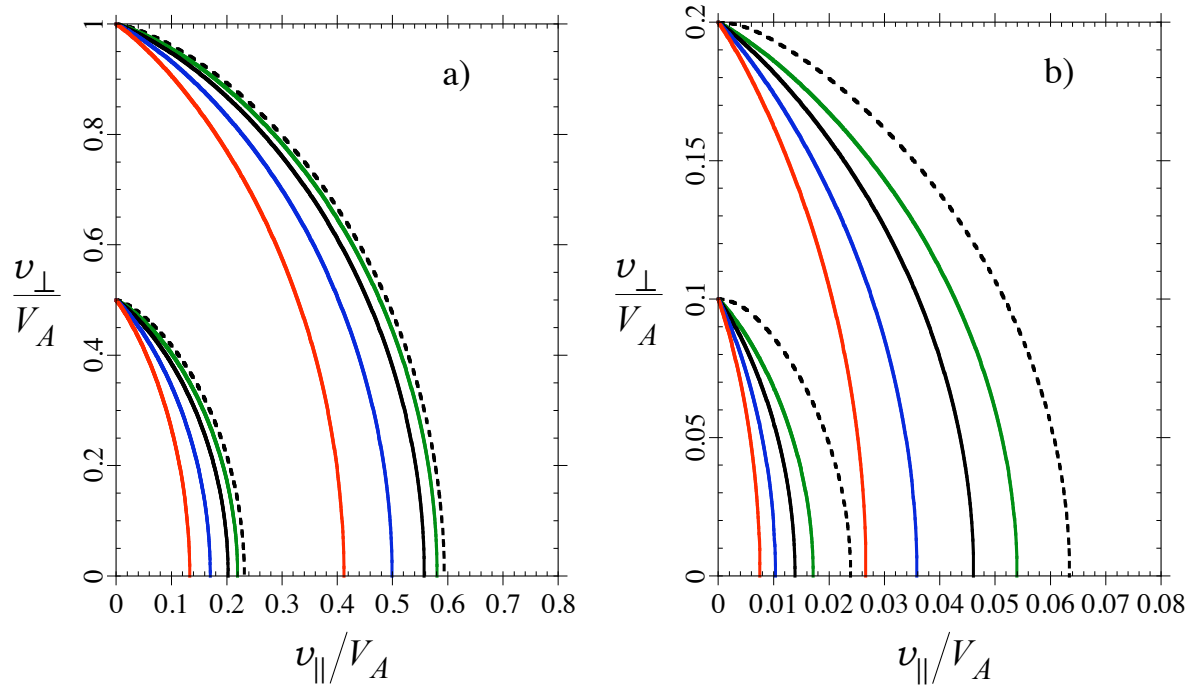

**Figure 4.** Representative resonant surfaces for protons in the marginally stable state. For each value of $\beta_\parallel$, we show surfaces for four values of $\eta$, equivalently their value of $\upsilon_\perp$ at $\upsilon_\parallel = 0$. a) $\eta = 0.5$ and 1.0. b) $\eta = 0.1$ and 0.2. The values of $\beta_\parallel$ and their color scheme are the same as in Figure 3: $\beta_\parallel = 0.001785$ (green), $\beta_\parallel = 0.01624$ (black), $\beta_\parallel = 0.1841$ (blue), and $\beta_\parallel = 1.694$. The black dashed lines show the equivalent surfaces obtained using the cold-plasma dispersion relation.

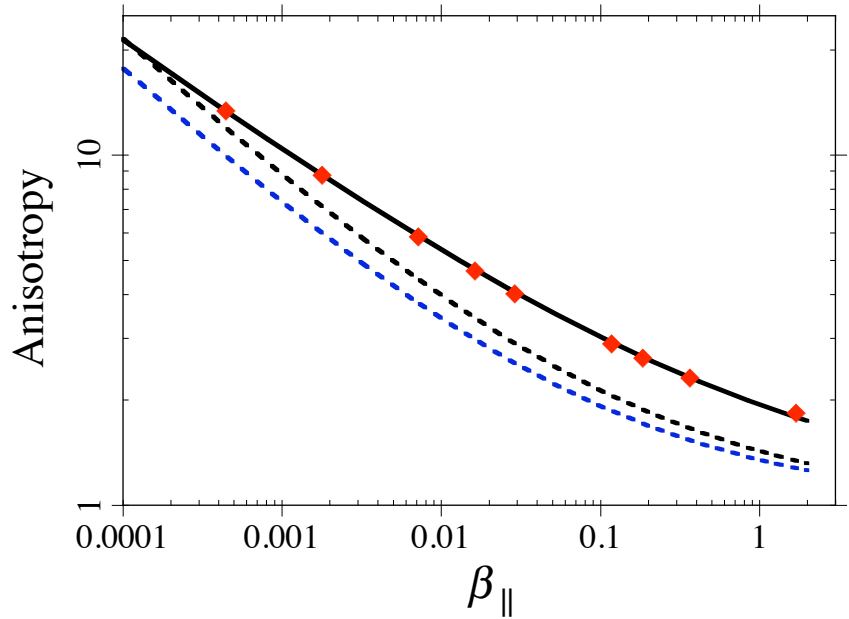

**Figure 5.** Proton anisotropies for the marginally stable state (red diamonds). For comparison, we show the bi-Maxwellian estimates from Gary et al (1994), for maximum growth rates $\gamma_{max} = 10^{-3}$ (black dashed line) and $10^{-4}$ (blue dashed line). The black solid line is our fit to the same functional form, with the parameter values given in (21).